\documentclass{my-mpla}

\usepackage{color}
\usepackage{amsmath,amssymb}
\usepackage{axodraw}
\newcommand{\be}{\begin{equation}}
\newcommand{\ee}{\end{equation}}
\newcommand{\bea}{\begin{eqnarray}}
\newcommand{\eea}{\end{eqnarray}}
\newcommand{\nn}{\nonumber}
\newcommand{\tw}{{\mbox{{\tiny W}}}}
\newcommand{\txred}{\textcolor{red}}

\newcommand{\Xsp}{X^{{\mbox{\tiny parton}}}_s}
\newcommand{\OO}[2]{(O_#1, O_#2)}

\begin{document}

\markboth{Andrea Ferroglia} {QCD Corrections to the Radiative
Decay $\bar{B} \to X_s \gamma$}

\catchline{}{}{}{}{}

\title{
QCD CORRECTIONS TO THE RADIATIVE DECAY $\bar{B} \to X_s \gamma$ }

\author{\footnotesize ANDREA FERROGLIA}

\address{Institute For Theoretical Physics, University of Z\"urich,
Winterthurerstrasse 190\\
Z\"urich, CH-8057, Switzerland \\
andrea.ferroglia@physik.uzh.ch}

\maketitle


\begin{abstract}
In this short review, the calculation of the
next-to-next-to-leading order  QCD corrections to the inclusive
radiative decay $\bar{B} \to X_s \gamma$ is described. I summarize
the salient features of the calculational framework adopted,
discuss the results obtained in the last few years, and indicate
the technical tools that made the NNLO calculations possible. I
conclude by comparing the current NNLO theoretical estimate for
the branching ratio with  the experimental measurement and by
briefly discussing the size and origin of the residual theoretical
uncertainty.

\keywords{B, meson, inclusive reaction, QCD, radiative correction.}
\end{abstract}

\ccode{PACS Nos.: 13.20.He,12.38.Bx.}

\section{Framework}

\begin{figure}[t]
\vspace*{1.15cm}
\[\vcenter{\hbox{
  \begin{picture}(0,0)(0,0)
\SetScale{1}
\SetColor{Orange}
  \SetWidth{1.5}
\ArrowLine(-70,15)(0,15)
  \SetWidth{.5}
\SetColor{Magenta}
\ArrowLine(0,15)(52,-10)
\SetColor{Black}
\ArrowLine(0,-15)(-70,-15)
\ArrowLine(44,-58)(0,-15)
\Photon(0,15)(50,65){3}{15}
\SetColor{Blue}
\Gluon(-60,15)(-40,-15){2}{10}
\Gluon(10,-25)(55,-35){2}{12}
\Gluon(-20,-15)(35,-2){2}{15}
\Gluon(8,11)(15,-5){2}{5.5}
\GlueArc(-30,15)(16,-180,-108.5){2}{5}
\GlueArc(-30,15)(16,-71.5,0){2}{5}
\ArrowArc(-30,-1)(5,0,180)
\ArrowArc(-30,-1)(5,180,360)
\Gluon(5,-20)(30,-15){2}{6}
\ArrowLine(30,-15)(45,-12)
\ArrowLine(37,-29)(30,-15)
\SetColor{Red}
\CCirc(0,15){3}{Black}{Green}
\SetColor{Black}
\COval(50,-35)(25,15)(0){Black}{Cyan}
\COval(-70,-0)(15,10)(0){Black}{Cyan}
\Text(52,-41)[cb]{{\Large $X_s$}} \Text(-71,-4)[cb]{$\bar{B}$}
\Text(55,50)[cb]{$\gamma$} \Text(-35,20)[cb]{$b$}
\Text(35,5)[cb]{$s$}
\end{picture}}}
\]
\vspace*{1.2cm} \caption{A schematic illustration of the $\bar{B}
\to X_s \gamma$ decay.\protect\label{fig1}}
\end{figure}
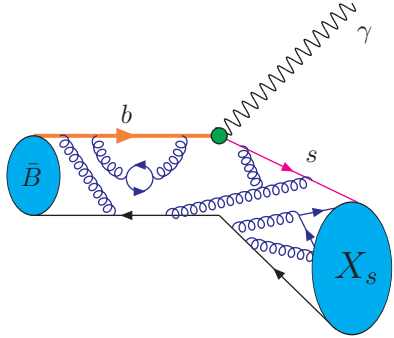

The inclusive radiative decays  of $B$ mesons into a photon and an
arbitrary hadronic state of total strangeness $-1$, $\bar{B} \to
X_s \gamma$,
 currently play a relevant role within the precision tests of the Standard Model
(SM) and of its extensions. The decay process is sketched in
Fig.~\ref{fig1}; $B$ denotes a $B^\pm$ or $B^0$  mesons, while
$X_s$ indicates an inclusive hadronic state not containing charmed
particles.

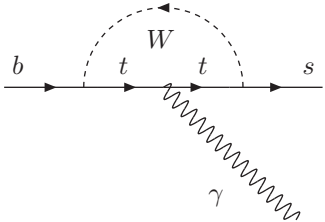
\begin{figure}[t]
\vspace*{.2cm}
\[\vcenter{\hbox{
  \begin{picture}(0,0)(0,0)
\SetScale{1}
  \SetWidth{.5}
\SetColor{Black}
\ArrowLine(-60,0)(-25,0) \ArrowLine(-25,0)(0,0)
\ArrowLine(0,0)(25,0) \ArrowLine(25,0)(60,0)
\DashArrowArc(0,0)(30,0,180){2} \Photon(0,0)(50,-50){3}{15.5}
\Text(-55,5)[cb]{$b$} \Text(55,5)[cb]{$s$} \Text(-1,15)[cb]{$W$}
\Text(-15,5)[cb]{$t$}
\Text(15,5)[cb]{$t$}\Text(20,-45)[cb]{$\gamma$}
\end{picture}}}
\]
\vspace*{.7cm} \caption{One of the penguin diagrams contributing
to $b \to s \gamma$ decay.
\protect\label{fig2}}
\end{figure}

 At the parton level, the process in Fig.~\ref{fig1} is
induced by a Flavor Changing Neutral Current (FCNC) decay of the
$b$ quark contained in the $\bar{B}$ meson. The $b$ quark decays
into a strange quark plus other partons, collectively indicated by
the symbol $\Xsp$, and a photon. In the SM, such a decay takes
place at first at one loop, through ``penguin'' diagrams such as
the one shown in Fig.~\ref{fig2}.
FCNC decays are rare (i.~e. loop-suppressed) in the SM; therefore,
they are  very sensitive to Beyond the SM physics effects, which
can arise in the perturbative expansion at the same order as the
leading SM contribution.

In contrast with the exclusive decay modes, inclusive decays of
$B$ mesons are theoretically clean observables; in fact, it is
possible to prove that the decay width $\Gamma (\bar{B} \to X_s
\gamma )$ is well approximated by the partonic decay rate $\Gamma
( b \to \Xsp \gamma )$:
\be \Gamma(\bar{B} \to X_s \gamma ) = \Gamma ( b \to \Xsp \gamma )
+ \Delta^{{\tiny {\rm non-pert.}}} \, . \label{HQE} \ee
The second term on the r.~h.~s. of Eq.~(\ref{HQE}) represents
non-perturbative corrections. The latter are small, since they are
suppressed at least by a factor $(\Lambda_{{\rm QCD}}/m_b)^2$,
where $m_b$ is the $b$-quark mass and $\Lambda_{{\rm QCD}} \sim
200$~MeV. The relation in Eq.~(\ref{HQE}) is known as {\em Heavy
Quark Expansion} (for a review, see Ref.~\refcite{ManoharWise}).

The partonic process can be studied within the context of perturbative QCD.
%
However, the first-order QCD corrections to the partonic process
are very large. The large corrections originate from hard gluon
exchanges between quark lines of the one-loop electroweak graphs
(see Fig.~\ref{fig3}). In general, Feynman diagrams involving
different mass scales (say $m_1$ and $m_2$, for example), depend
on logarithms of the ratio of the masses. If there is a strong
hierarchy among the two mass scales (e.g. $m_1 \ll m_2$), then the
logarithms are numerically large. In the case of QCD corrections
to the partonic process $b \to \Xsp \gamma$, the mass scales
involved are the $W$-boson mass $M_\tw$, the top quark mass $m_t$,
and $m_b$. $M_\tw$ and $m_t$ are of the same order of magnitude
$\mu_\tw \sim 100$~GeV, while the $b$-quark mass is considerably
smaller: $m_b \sim 5$~GeV. At $n$-th order in the perturbative
 expansion in the  strong coupling constant $\alpha_s (m_b)$, one finds terms
 of the form
\be
\alpha^n_s(m_b) \ln^m \left( \frac{m_b}{\mu_\tw}\right) \, ,
\label{LLogs}
\ee
where $m \leq n$.  For $n=1$ and $m=1$, the product in
Eq.~(\ref{LLogs}) is too large to be used as an expansion
parameter; terms enhanced by large logarithmic coefficients must
be resummed at all orders. Conventionally, calculations performed
by resumming logarithms in which $m=n$ are referred to as {\em
leading order} (LO) precision calculations. By resumming terms in
which $m=n,n-1$, it is possible to obtain results of {\em
next-to-leading order} (NLO) precision. Similarly, resumming
$m=n,n-1,n-2$ logarithms, it is possible to achieve {\em
next-to-next-to-leading order} (NNLO) precision.

The easiest  way to implement the resummation of the large
logarithms  discussed above is to work within the context of a
renormalization-group-improved effective theory with five active
quarks. In such a theory, the heavy degrees of freedom involved in
the decay under study are integrated out. By means of an operator
product expansion, it is possible to factorize the contribution of
the short-distance and long-distance dynamics in the decay of the
$B$ meson. In the SM, the short-distance dynamic is characterized
by mass scales of  the order of the top-quark or $W$-boson mass,
while the long-distance dynamic is characterized by the $b$-quark
mass. The boundary between short-distance and long-distance is
chosen at a low-energy scale $\mu_b$ such that $m_b \sim \mu_b \ll
M_\tw$. Clearly, the  scale $\mu_b$ is unphysical, and therefore
physical quantities should not depend on it. However, all
calculations are performed up to some fixed order in perturbation
theory. This truncation of the perturbative series induces a
dependence on the low-energy scale in
 the physical observables,
which is formally of higher order with respect to the precision goal
of the calculation.
%
%
 From  the practical point of view, the Lagrangian  employed in calculating the $b \to \Xsp \gamma$
 decay rate can be written
as\footnote{Terms suppressed by the CKM factor $V^*_{us} V_{ub}$
are ignored; however, their NLO effect was accounted for in the
NNLO calculation of the $\bar{B} \to X_s \gamma$ branching ratio
in Ref.~\refcite{MisSte}.}
\be
{\mathcal L} = {\mathcal L}_{\mbox{QED}\otimes\mbox{QCD}}\left(\mbox{u,d,c,s,b}\right)
+\sum_{i = 1}^{8} \frac{4 G_F}{\sqrt{2}}V_{ts}^* V_{tb}
 C_i(\mu,\mu_\tw) O_i(\mu) + {\mathcal O}\left(\frac{m_b}{M_\tw} \right)\, .
\label{Lag}
\ee
In the equation above,  ${\mathcal L}_{\mbox{QED}}$  and
${\mathcal L}_{\mbox{QCD}}$
 represent the usual QED and QCD Lagrangians with five active quark flavors.
 The second term in Eq.~(\ref{Lag}) is more interesting: $G_F$ is the Fermi
 constant, $V_{ts}$ and $V_{tb}$ are elements of the CKM matrix,
and $O_i$ are eight {\em effective operators} of dimensions five
and six. Operators with dimensions larger than six are suppressed
by inverse powers of the $W$-boson mass and are ignored. Finally,
the short-distance dynamic is encoded in the ``coupling
constants'' of the effective operators, which are called {\em
Wilson coefficients} and are indicated by $C_i$ in
Eq.~(\ref{Lag}). The Wilson coefficients are the only elements of
the Lagrangian which depend on the heavy particles masses
$M_\tw$ and $m_t$.
%

Any perturbative calculation of the $b \to \Xsp \gamma$ decay rate
within the context of the renormalization-group-improved
perturbation theory applied to
the Lagrangian in Eq.~(\ref{Lag}) requires three different steps:
\begin{itemize}
\item[i)] The first step, conventionally called {\em matching}, consists in
fixing the value of the Wilson coefficients at the high-energy
scale $\mu_\tw \sim M_\tw, m_t$. This is achieved by requiring
that Green functions calculated in the full SM and in the
effective theory provide the same result up to terms suppressed by
the ratio between the external momenta and $M_\tw$ or $m_t$.  At
the scale $\mu_\tw$, QCD corrections are small and subsequently
can be calculated in fixed-order perturbation theory.

\item[ii)] Secondly, once the value of the Wilson coefficient at the
electroweak scale has been obtained from the matching step, it is then
 necessary to
obtain the value of the Wilson coefficients at the low-energy
scale $\mu_b \sim m_b$. This can be achieved by solving the
system of renormalization group equations (RGE) satisfied by the
Wilson coefficient, which have the following form:
\be \mu \frac{d}{d \mu} C_i(\mu) = \gamma_{ji}(\mu) C_j(\mu) \, .
\label{RGE} \ee
The matrix $\gamma$ in the equation above is the anomalous
dimension matrix (ADM) of the effective operators. Since the
various operators mix under renormalization, this step of the
calculation is called {\em mixing}. By solving the RGE, it is
possible to resum at all orders
 the large logarithms of the ratio $\mu_\tw/\mu_b$  in the Wilson coefficients.
This becomes evident by looking at the leading order solution of
Eq.~(\ref{RGE}) (for the diagonalized ADM, indicated by a tilde)
\be \tilde{C}_i(\mu_b) = \left[1 + \beta_0
\frac{\alpha_s(\mu_b)}{4 \pi} \ln\left(\frac{\mu_\tw}{\mu_b}
\right) \right]^{-\tilde{\gamma}^{(0)}_{ii}/2 \beta_0} \tilde{C}_i
(\mu_\tw) \, , \ee
where $\beta_0= 11 -2/3 N_f$ is the leading order QCD beta
function, and $N_f$ is the number of active quarks.

\item[iii)] Finally, it is necessary to calculate on-shell matrix elements of
the partonic process in the effective theory.
QCD radiative corrections to the matrix elements do not
include large logarithms, since the dependence on the heavy degrees
of freedom is completely encoded within the Wilson coefficients.
\end{itemize}

\begin{figure}[t]
\vspace*{.1cm}
\[\vcenter{\hbox{\begin{picture}(0,0)(0,0)
\SetScale{1}
  \SetWidth{.5}
\SetColor{Black}
\ArrowLine(-60,0)(-25,0) \ArrowLine(-25,0)(0,0)
\ArrowLine(0,0)(25,0) \ArrowLine(25,0)(60,0)
\DashArrowArc(0,0)(30,0,180){2} \Photon(0,0)(50,-50){3}{15.5}
\SetColor{Red} \GlueArc(35,0)(16,-180,0){2}{12}
\Text(-55,4)[cb]{$b$}\Text(55,4)[cb]{$s$}\Text(-20,4)[cb]{$t$}\Text(20,4)[cb]{$t$}
\Text(-1,16)[cb]{$W$}
\Text(50,-25)[cb]{$g$}\Text(20,-45)[cb]{$\gamma$}
\end{picture}}}
\]
\vspace*{.7cm} \caption{One of the diagrams contributing to the
first order QCD corrections to $b \to s \gamma$
decay.\protect\label{fig3}}
\end{figure}

The eight effective operators appearing in the  Lagrangian in
Eq.~(\ref{Lag}) are listed in \ref{Operators}.
At LO, the $b \to s \gamma$ amplitude is proportional to the {\em
effective} Wilson coefficient $C_7^{\mbox{{\tiny eff}}}$; the
effective coefficients were introduced in
Ref.~\refcite{Buras:1993xp}.


Radiative decays of the $B$ meson were first experimentally
observed (in the exclusive $B \to K^* \gamma$ decay mode) by the
CLEO collaboration at Cornell in 1993. Nowadays, the branching
ratio of the inclusive decay $\bar{B} \to X_s \gamma$ has been
measured by several collaborations. The current world average,
obtained by averaging the CLEO, BELLE, and BABAR measurements
(Refs.~\refcite{CLEO,BELLE,BABAR1,BABAR2,BABAR3}), is
(Ref.~\refcite{worldav})
\be {\mathcal B}\left(\bar{B} \to X_s\,
\gamma\right)^{{\mbox{{\tiny WA}}}}_{E_\gamma > E_0} = (3.52 \pm
0.23 \pm 0.09) \times 10^{-4} \, . \label{WA} \ee
In Eq.~(\ref{WA}) the first error is due to statistical and
systematic uncertainty, while the second is due to theory input on
the $b$-quark Fermi motion. In order to eliminate irreducible
backgrounds, experimental collaborations impose a lower cut on the
photon energy.
The value in Eq.~(\ref{WA}) refers to a lower cut $E_0 = 1.6$~GeV.
In view of current experimental accuracy, theoretical predictions
based upon NLO calculations in the
$\alpha_s$ expansion
 are no longer sufficient.
In the last eight years, partial results necessary for the
theoretical calculation of the $\bar{B} \to X_s \gamma$ branching
ratio with NNLO precision were obtained by several groups
(Refs.~\refcite{NLOc4,Misiak:2004ew,Gorbahn:2004my,Gorbahn:2005sa,Czakon:2006ss,Bieri:2003ue,Blokland:2005uk,Asatrian:2006ph,Ligeti:1999ea,Melnikov:2005bx,Asatrian:2006sm,MisSte}).
 Two years ago, by employing these  partial results,
it was possible to obtain the first theoretical estimate of the
branching ratio at NNLO (\refcite{MisSte,manyaut}). Currently, the
NNLO program is heading toward completion
(Ref.~\refcite{Asatrian:2006rq,Ewerth:2008nv,Boughezal:2007ny}).

In order to calculate the $\bar{B} \to X_s \gamma$ branching ratio
at NNLO, it is necessary to make use of some of the most powerful
and recent techniques available for multi-loop Feynman diagrams
calculations. The aim of the rest of this short review is to
indicate the techniques employed in the parts of the NNLO
calculation carried out so far, as well as to indicate the areas
in which work is currently in progress. I conclude by comparing
the current theoretical prediction of the branching ratio with the
experimental measurements and by briefly discussing the residual
theoretical uncertainties.

%
%
%
%
%
%
%

\section{NNLO Calculation}

The experimental error on the $\bar{B} \to X_s \gamma$ branching
ratio in Eq.~(\ref{WA}) is of $7 \%$.
 The theoretical error
affecting the NLO prediction  is about $10 \%$ (see for example
the comprehensive reviews in
Refs.~\refcite{Buras:2002er,Hurthrev}). Aside for the NLO QCD
corrections
(Refs.~\refcite{NLOc,NLOc2,NLOc3,NLOc5,NLOb,NLOb2,NLOa,NLOa2,NLOa3,NLOa4,NLOa5,NLOa6,NLOa7}),
NLO determinations of the branching ratio include electroweak
effects
(Refs.~\refcite{Czarnecki:1998tn,Baranowski:1999tq,Gambino:2000fz});
moreover, the numerical impact of the charm- and bottom-quark
masses was analyzed in detail in Ref.~\refcite{Gambino:2001ew}.
However, the NNLO QCD corrections were estimated to be  at the
level of $\pm 7 \%$ already in Ref.~\refcite{Kagan:1998ym}. The
inclusion of the NNLO QCD corrections is expected to significantly
reduce the uncertainty associated to the charm-quark mass
renormalization scale. In fact, since the charm-quark mass $m_c$
first enters the branching ratio
 at NLO,  the related scale dependence is a NNLO issue (see
Ref.~\refcite{Gambino:2001ew}). It is thus mandatory to include
the numerically leading NNLO QCD effects in the theoretical
prediction of the $\bar{B} \to X_s \gamma$ branching ratio in
order to bring the theoretical uncertainty at the same level of
the experimental one. Below matching, mixing, and matrix element
calculation at NNLO in QCD are discussed.

\subsection{Matching}

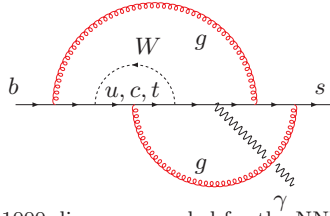
\begin{figure}[t]
\vspace*{.1cm}
\[\vcenter{\hbox{\begin{picture}(0,0)(0,0)
\SetScale{0.5}
  \SetWidth{.5}
\SetColor{Black}
\ArrowLine(-120,0)(-90,0) \ArrowLine(-90,0)(-60,0)
\ArrowLine(-60,0)(-30,0) \ArrowLine(-30,0)(0,0)
\ArrowLine(0,0)(30,0) \ArrowLine(30,0)(60,0)
\ArrowLine(60,0)(90,0) \ArrowLine(90,0)(120,0)
\DashArrowArc(-30,0)(30,0,180){2} \Photon(30,0)(69,-39){3}{11}
\Photon(76,-46)(90,-60){3}{4} \SetColor{Red}
\GlueArc(30,0)(60,-180,0){2}{42} \GlueArc(-15,0)(75,0,180){2}{50}
\Text(-61,4)[cb]{{\small $b$}}\Text(55,4)[cb]{{\small $s$}}
\Text(-16,1)[cb]{{\small $u,c,t$}} \Text(-10,19)[cb]{{\small $W$}}
\Text(10,-26)[cb]{{\small $g$}}\Text(40,-40)[cb]{{\small
$\gamma$}}\Text(10,21)[cb]{{\small $g$}}
\end{picture}}}
\]
\vspace*{.3cm} \caption{One of approximately  1000 diagrams needed
for the NNLO matching of the $O_7$ operator.\protect\label{figMathcing}}
\end{figure}

The goal of the matching procedure at NNLO is to calculate the
order $\alpha_s^2$ corrections to the Wilson coefficients
evaluated at the high-energy scale $\mu_\tw \sim m_t, M_\tw$. The
matching of the four-quark operators $O_1, \cdots,O_6$, which
requires the calculation of two-loop Feynman diagrams in the SM,
was carried out in Ref.~\refcite{NLOc4}. Since the magnetic- and
chromo-magnetic dipole operators $O_7$ and $O_8$ first arise  at
one-loop in the SM, the matching of these operators requires
 the calculation of
three-loop Feynman diagrams, one of which is shown in
Fig.~\ref{figMathcing}. The dipole operator matching was carried
out in Ref.~\refcite{Misiak:2004ew}. Both Ref.~\refcite{NLOc4} and
Ref.~\refcite{Misiak:2004ew} employ the same calculational
technique. The SM diagrams are expanded in the ratio
$(\mbox{external momenta})^2/\mu_\tw^2$ up to the first
non-vanishing order. Spurious infrared divergencies originating
from the expansion are regulated in dimensional regularization.
%
%
 All the masses, with the exception of $M_\tw$, $m_t$, and $m_b$,
are set equal to zero from the start; terms proportional to
$m_b^2$ are also neglected. The integrals that must be evaluated
after the expansion in the external momenta correspond to vacuum
diagrams, involving one or, as in Ref.~\refcite{Misiak:2004ew},
two mass scales. Two- and three-loop vacuum integrals depending on
a single mass scale are known. Three-loop vacuum diagrams
depending on two different mass scales could not be evaluated
directly. The problem was solved by expanding the integrals around
the point $m_t= M_\tw$ and for $m_t \gg M_\tw$.  Both expansions
give satisfactory results for the physical value $M_\tw \sim
m_t/2$. In order to write down the matching equations, it is
necessary to require equivalency of the SM and effective theory
Green functions. The calculation of the latter is not problematic;
in fact, after expanding in the external momenta, loop diagrams
give rise to scaleless integrals, which vanish in dimensional
regularization.


\subsection{Mixing}
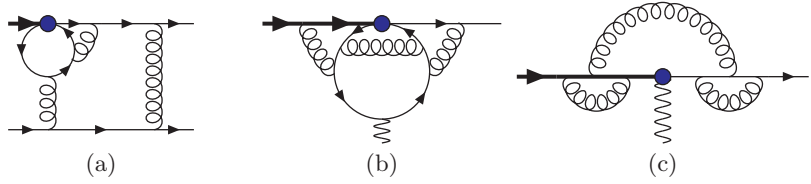
\begin{figure}[t]
\vspace*{.1cm}
\[\vcenter{\hbox{
  \begin{picture}(0,0)(0,0)
\SetScale{1} \SetWidth{1.5}\ArrowLine(-35,20)(-20,20)
  \SetWidth{.5}
\SetColor{Black}
 \ArrowLine(-20,20)(0,20)
\ArrowLine(0,20)(20,20) \ArrowLine(20,20)(35,20)
\ArrowLine(-35,-20)(-20,-20) \ArrowLine(-20,-20)(20,-20)
\ArrowLine(20,-20)(35,-20) \Gluon(20,20)(20,-20){3}{8.5}
\Gluon(-20,0)(-20,-20){3}{3.5}
\ArrowArc(-20,10)(10,90,270)\ArrowArc(-20,10)(10,-90,0)
\ArrowArc(-20,10)(10,0,90) \CCirc(-20,20){3}{Black}{Blue}
\GlueArc(-20,20)(16,-56,0){3}{2.5} \Text(0,-38)[cb]{{\small (a)}}
\end{picture}
\hspace*{3.5cm}
  \begin{picture}(0,0)(0,0)
\SetScale{1} \SetWidth{1.5}
\ArrowLine(-45,20)(-30,20)\ArrowLine(-30,20)(0,20)
  \SetWidth{.5}
\SetColor{Black}
\ArrowLine(0,20)(45,20) \ArrowArc(0,2)(18,90,155)
\ArrowArc(0,2)(18,155,270)
\ArrowArc(0,2)(18,-90,25)\ArrowArc(0,2)(18,25,90)
\GlueArc(10,30)(40,194,226){3}{4.5}
\GlueArc(-10,30)(40,-46,-14){3}{4.5}\Photon(0,-16)(0,-25){3}{2.5}
\Gluon(15.5,11)(-15.5,11){3}{6.5}
\CCirc(0,20){3}{Black}{Blue}\Text(0,-38)[cb]{{\small (b)}}
\end{picture}
\hspace*{3.5cm}
  \begin{picture}(0,0)(0,0)
\SetScale{1} \SetWidth{1.5}
\ArrowLine(-55,0)(-40,0)\Line(-40,0)(0,0)
  \SetWidth{.5}
\SetColor{Black}
\Line(0,0)(40,0) \ArrowLine(40,0)(55,0)
\GlueArc(0,0)(25,0,180){3}{14.5} \GlueArc(-25,0)(10,180,360){3}{6}
\GlueArc(25,0)(10,180,360){3}{6} \Photon(0,0)(0,-25){3}{5.5}
\CCirc(0,0){3}{Black}{Blue}\Text(0,-38)[cb]{{\small (c)}}
\end{picture}
}}
\]
\vspace*{.4cm} \caption{Examples of Feynman diagrams needed for
the calculation of the matrices $A^{(2)}$ (a), $B^{(2)}$ (b), and
$C^{(2)}$ (c). Thick arrow lines indicate $b$ quarks, thin arrow
lines indicate massless quarks; the dark dots indicate effective
operators.\protect\label{fig8}}
\end{figure}

The entries of the ADM $\gamma$ in Eq.~(\ref{RGE}) can be obtained
from the QCD renormalization constants in the effective theory.
The latter are derived from ultraviolet divergencies in the
Feynman diagrams with effective operator insertions. The ADM has
the following perturbative expansion
\be \gamma(\mu) = \sum_{k=0} \left( \frac{\alpha_s(\mu)}{4
\pi}\right)^{(k+1)} \gamma^{(k)} \, , \label{ADMexp} \ee
where the matrices at each order in $\alpha_s$ have a block  structure:
\be
 \gamma^{(k)} = \left( \begin{array}{cc}
  A^{(k)}_{6\times6} &B^{(k)}_{6\times 2}\\
  0 _{2\times6}  &  C^{(k)}_{2\times2}
   \end{array}\right) \, .
\ee
The lower-left block vanishes because the dimension-five dipole
operators do not generate UV divergencies in dimension-six,
four-quark amplitudes. The matrices $A^{(k)}$ and $C^{(k)}$ can be
obtained from the UV divergencies of $(k+1)$-loop diagrams, while
to evaluate $B^{(k)}$ it is necessary to evaluate the UV
divergencies of $(k+2)$-loop diagrams. The matrices $A^{(2)}$,
$B^{(2)}$, and $C^{(2)}$  were calculated in the
Refs.~\refcite{Gorbahn:2004my,Czakon:2006ss}, and
\refcite{Gorbahn:2005sa}, respectively. Sample diagrams needed in
the calculation of the NNLO ADM are shown in Fig.~\ref{fig8}.
%
Since one is interested in Feynman diagrams UV divergencies, it is
necessary to evaluate the corresponding integrals by regulating UV
and infrared singularities in different ways. In
Refs.~\refcite{Gorbahn:2004my,Czakon:2006ss},and
\refcite{Gorbahn:2005sa} this was done by introducing a common
mass $M$ for all fields,  and thus by expanding the loop integrals
in inverse powers of $M$. Therefore, the only integrals needed
were single scale tadpoles up to four loops;  these integrals are
known. The number of diagrams involved is impressive: the
calculation of $B^{(2)}$ alone requires the evaluation of over
20000 four-loop Feynman graphs.

\subsection{Matrix Elements - Total Rate}

To complete the calculation of the NNLO corrections to the $b \to
\Xsp \gamma$ decay rate, it is necessary to calculate the
$O(\alpha_s^2)$ corrections to the matrix elements in the
low-energy effective theory of Eq.~(\ref{Lag}). The decay width
for the process\footnote{I consider  squared amplitudes summed
over  spin, color, and polarization of the final state, as well as
averaged over the spin and color of the incoming $b$ quark. } $b
\to \Xsp \gamma$ can be written as
\be
\Gamma(b\to \Xsp \gamma)_{E_\gamma>E_0} =\!
\frac{G_F^2\alpha_{\rm
 em}\bar{m}_b^2(\mu)m_b^3}{32\pi^4}\!|V_{tb}^{}V_{ts}^*|^2\!
\sum_{i,j}C_{i}^{\rm eff}(\mu)\,C_{j}^{\rm
eff}(\mu)\,G_{ij}(E_0,\mu) \, . \label{decay}
\ee
The term proportional to $G_{ij}$ in Eq.~(\ref{decay}) originates
from the interference of diagrams mediated by the effective
operator $O_i$ and diagrams involving the effective operator
$O_j$. Consequently, $G_{ij}$ is referred to as the $\OO{i}{j}$
component of the decay width. In Eq.~(\ref{decay}) $m_b$ and
$\bar{m}_b$ indicate the pole and running $\overline{\mbox{MS}}$
bottom quark mass, respectively. $\alpha_{em}$ is the fine
structure constant at zero momentum transfer. The total decay rate
can be obtained from Eq.~(\ref{decay}) by setting the lower cut on
the photon energy, $E_0$, equal to zero. In this section we
discuss the calculation of NNLO QCD corrections to the total decay
rate.

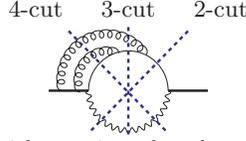
\begin{figure}[t]
\vspace*{.1cm}
\[\vcenter{\hbox{\begin{picture}(0,0)(0,0)
\SetScale{.5}
  \SetWidth{1.8}
\Line(-60,0)(-30,0)
\Line(60,0)(30,0)
  \SetWidth{.5}
\CArc(0,0)(30,0,180)
\PhotonArc(0,0)(30,-180,0){2}{16.5}
%
%
\GlueArc(-16,8)(35,33,193){3}{14}
\GlueArc(-16,8)(22,67,204){3}{7}

\Text(0,28)[cb]{{\small 3-cut}}
\Text(-35,28)[cb]{{\small 4-cut}}
\Text(35,28)[cb]{{\small 2-cut}}

%
%
  \SetWidth{1.8}
\SetColor{Blue}
\DashLine(-45,45)(28,-30){3}
\DashLine(45,45)(-28,-30){3}
\DashLine(0,48)(0,-33){3}
\end{picture}
}}\] \caption{Two-, three-, and four-particle cuts  in a
three-loop $b$-quark self-energy diagram; they all contribute to
the $\OO{7}{7}$ component of the NNLO $b \to \Xsp g$ decay rate.
Thick lines indicate $b$ quarks, thin lines indicate strange
(massless) quarks. \protect\label{fig7}}
\end{figure}

The $\OO{7}{7}$ component is the only one completely known at NNLO in QCD.
This component is numerically dominant  and was independently
calculated by two different groups
(Refs.~\refcite{Blokland:2005uk,Asatrian:2006ph}). In order to
calculate the NNLO QCD corrections to  the $\OO{7}{7}$ component
of the decay width in Eq.~(\ref{decay}), it is necessary to
consider three different sets of matrix elements:
\begin{itemize}
\item Two-loop corrections to the process $b \to s \gamma$ interfered with
the tree-level matrix element of the $b \to s \gamma$ decay,
\item one-loop corrections to the process $b \to s \gamma g$ interfered with
the corresponding tree-level amplitude, and
\item tree-level matrix elements for the processes $b \to s \gamma g g$
and $b \to s \gamma q \bar{q}$ interfered among themselves.
\end{itemize}
The contribution of each of the quantities listed above to the $b
\to \Xsp \gamma$ decay width is directly related to the imaginary
part of three-loop $b$-quark self-energy diagrams  by means of the
optical theorem (an example is shown in Fig.~\ref{fig7}). The
contribution of each physical cut to the imaginary part of a self
energy can be evaluated by means of the Cutkosky rules
(Refs.~\refcite{cut1,cut2,cut3}).
In Ref.~\refcite{Asatrian:2006ph}, the contribution of each cut of
each three-loop $b$-quark self-energy diagram was evaluated  by
means of the methods commonly employed in the calculation of
multi-loop Feynman diagrams.
%
By employing the Laporta algorithm
(Refs.~\refcite{La1,La2,La3,La4}),
 it is possible to rewrite each  cut diagram as a linear combination of
a small number of Master Integrals (MIs). The Cutkosky rules
associate  a Dirac delta function with each line going through the
cut. The latter sets the momentum flowing through a  ``cut
propagator'' on its mass-shell. The delta functions can be written
as a difference of propagators (Refs.~\refcite{cp1,cp2}); if $q$
is the momentum flowing through a cut propagator one finds
\be
\delta\left( q^2 + m^2\right) = -\frac{1}{2 \pi i}
\left(\frac{1}{q^2 + m^2 + i 0} - \frac{1}{q^2 +m^2 -i 0} \right) \, .
\label{cutprop}
\ee
The reduction procedure is simplified by the fact  that all of the
integrals in which one of the cut propagators vanishes or is
raised to a negative power are equal to zero, since in those cases
the $\pm i 0$ prescription in Eq.~(\ref{cutprop}) becomes
irrelevant. In Ref.~\refcite{Asatrian:2006ph}, the MIs were
evaluated numerically by means of the sector decomposition method
(Ref.~\refcite{sd1}); the four-particle phase space integrals were
parameterized according to the methods presented in
Refs.~\refcite{PSpar1} and \refcite{PSpar2}. After summing over
the contribution of all of the cuts, the residual poles in the
dimensional regulator $\varepsilon$ are canceled by UV
renormalization. The methods employed in
Ref.~\refcite{Asatrian:2006ph} can also be applied to the
calculation of the $\OO{7}{8}$ and  $\OO{8}{8}$ components of the
decay width in Eq.~(\ref{decay}).
%
The authors  of Ref.~\refcite{Blokland:2005uk} were able to
analytically calculate the $\OO{7}{7}$ component of the $b \to
\Xsp \gamma$ decay width. In fact, in
Ref.~\refcite{Blokland:2005vq}, the same authors evaluated the
complete imaginary part of individual three-loop $b$-quark self
energy diagrams. Each imaginary part is the sum of all of the
possible cuts present in a given graph. In the case of the
$\OO{7}{8}$ and $\OO{8}{8}$ decay width components, not all of the
cuts correspond to the interference of $b \to \Xsp \gamma$ matrix
elements. Therefore, in order to calculate the $\OO{7}{8}$ and
$\OO{8}{8}$ components, it will be necessary to individually
evaluate  each cut contributing to the process.

The first calculation of $b \to \Xsp \gamma$ matrix elements at
NNLO was the one described in Ref.~\refcite{Bieri:2003ue}.  In
that work,  a set of NNLO virtual corrections to the matrix
elements involving the effective operators\footnote{At NNLO it is
possible to neglect the operators $O_3,\cdots,O_6$, since they are
suppressed by small Wilson coefficients; the NLO contribution to
the branching ratio arising from these operators is $< 1 \%$. }
$O_1$,$O_2$, $O_7$, $O_8$, as well as bremsstrahlung corrections
to matrix elements including $ O_7$, are evaluated. The authors
restrict their calculation to diagrams with a closed light-quark
loop. In fact, once the correction proportional to $\alpha^2_s
N_l$ (where $N_l$ is the number of light quarks) are known, it is
possible to estimate the complete NNLO corrections by means of the
naive non-abelianization hypothesis (NNA, see
Refs.~\refcite{Brodsky:1982gc,Beneke:1994qe}): the BLM (or
large-$\beta_0$) approximation is derived by replacing the factor
$N_l$ with the coefficient $-3 \beta_0 /2$  in diagrams with a
light-quark loop. In Ref.~\refcite{Bieri:2003ue}, the calculation
was carried out by means of Feynman parameterization of the
integrands, followed by an integration technique based upon
Mellin-Barnes (MB) transform (see Refs.~\refcite{MB1,MB2,MB3}).
The results are presented in analytic form. The three-loop virtual
corrections to the $O_1$ and $O_2$ matrix elements are expanded in
powers of $m_c^2/m_b^2$.

In Ref.~\refcite{MisSte}, Misiak and Steinhauser computed the NNLO
corrections in the BLM approximation in the non-physical limit
$m_c \gg m_b/2$. They observed that their result, evaluated at
$m_c \sim m_b/2$,  matches well with the calculation of the same
set of corrections in the small $m_c$ expansion
(Ref.~\refcite{Bieri:2003ue}). Because of this match, Misiak and
Steinhauser also evaluated the complete $m_c$-dependent NNLO
corrections to the $b \to \Xsp \gamma$ matrix elements in the $m_c
\gg m_b/2$ approximation. Subsequently, they assumed that the BLM
result is a good approximation of the full NNLO corrections for
$m_c =0$, and they interpolated their results for the non-BLM
corrections down to the measured value of $m_c$. The calculational
technique employed in Ref.~\refcite{MisSte} is the same technique
employed in the three-loop Wilson coefficient calculation
presented in Ref.~\refcite{Misiak:2004ew}.
%
The results of Refs.~\refcite{Bieri:2003ue} and \refcite{MisSte}
were crucial to obtain the first NNLO estimate of the $\bar{B} \to
X_s \gamma$ branching ratio (Refs~\refcite{MisSte,manyaut}).
However, the calculation of the NNLO QCD corrections to the $b \to
\Xsp \gamma$ matrix elements is not yet complete, and several
groups are still working to improve the understanding of this set
of corrections.

In Ref.~\refcite{Boughezal:2007ny}, the authors reconsider the
virtual fermion loop corrections to the matrix elements involving
the operators $O_1$ and $O_2$. They check the result of
Ref.~\refcite{Bieri:2003ue} by evaluating the corrections
proportional to $\alpha_s^2 N_l$, and they also numerically
calculate the diagrams involving a closed massive charm or bottom
quark loop. In the case of diagrams involving massless quark loops
the reduction is carried out by means of the Laporta algorithm,
while the MIs are evaluated in two different ways; first in an
 expansion in powers of $m_c/m_b$, and then by the numerical
 evaluation of the MB representation of the integrals.
The Laporta algorithm is also employed for diagrams involving a massive quark
loop. However, for some of the MIs
 encountered in the latter case, the numerical
 evaluation of MB  representations is not sufficiently precise. In these
 cases, an interesting technique based upon the numerical solution of the system
 of differential equations satisfied by the MIs is adopted.

The effects of the charm-quark mass in the $\OO{7}{7}$ and $\OO{7}{8}$
decay width components are analyzed in Ref.~\refcite{Asatrian:2006rq}
and Ref.~\refcite{Ewerth:2008nv}.

Currently, a group is working on the numerical calculation of the
$\OO{1}{7}$ and $\OO{2}{7}$ components in the  $m_c =0$
approximation
(Refs.~\refcite{Boughezal:2007km,Schutzmeier:2008sm}). The latter
represents the first step toward the calculation of the complete
set of NNLO corrections in the $m_c =0$ limit. Once this result becomes
available, it will then  be possible to interpolate the results of
Ref.~\refcite{MisSte} to the physical value of $m_c$ without
making any assumption on the behavior of the corrections for
vanishing $m_c$. In turn, this procedure will reduce the error
associated with the $m_c$ interpolation.

\subsection{Matrix Elements - Spectrum}
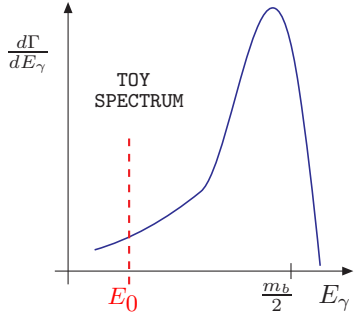
\begin{figure}[t]
\vspace*{2.5cm}
\[\vcenter{\hbox{
  \begin{picture}(0,0)(0,0)
\SetScale{1}
  \SetWidth{.4}
\SetOffset(-50,0)
\LongArrow(-5,0)(106,0) \LongArrow(0,-5)(0,100) \Line(84,3)(84,-3)
\Text(79,-15)[cb]{$\frac{m_b}{2}$}
\Text(27,70)[cb]{{\footnotesize{\tt TOY }}}
\Text(27,62)[cb]{{\footnotesize{\tt SPECTRUM}}}
  \SetWidth{.6}
\Text(101,-15)[cb]{$E_\gamma$}
\Text(-15,75)[cb]{$\frac{d \Gamma}{d E_\gamma}$}
  \SetColor{Blue}
\Curve{(10,8)(50,30)(87,69)(95,2)}
  \SetWidth{.8}
 \SetColor{Red}
\DashLine(23,50)(23,-5){3}
\Text(21,-15)[cb]{{\small \txred{$E_{\mbox{0}}$}}}
\end{picture}}}
\]
\vspace*{-.3cm} \caption{A sketch of the photon energy spectrum in
the $\bar{B} \to X_s \gamma$ decay. Only photons with energy
$E_\gamma > E_0$ are observed. \protect\label{fig5}}
\end{figure}

In the two-body $b \to s \gamma$ decay, the energy of the outgoing
photon is fixed by the kinematic of the process: $E_\gamma =
m_b/2$ in the $b$-quark rest frame. However, two different
phenomena give raise to a photon energy spectrum in the $\bar{B}
\to X_s \gamma$ decay. First, the partonic decay in which one is
interested is not $b \to s \gamma$ but rather $b \to \Xsp \gamma$;
in events involving the emission of one or more gluon or $q
\bar{q}$ pair, the photon has an energy $E_\gamma < m_b/2$.
Secondly, the Fermi motion of the $b$ quark in the $B$ meson also
contributes in generating a non-trivial photon energy spectrum.
The situation is sketched in Fig~\ref{fig5}; the smearing of the
photon energy spectrum beyond the partonic end point $m_b/2$ is
due to the non-perturbative Fermi motion effects, while the long
low-energy tail has its origin in the gluon/quark-pair
bremsstrahlung. The latter can be studied in perturbative QCD,
while Fermi motion effects are modeled by means of a
process-independent shape function  (see
Ref.~\refcite{Kagan:1998ym,Ligeti:2008ac}).
%
%
Since experimental collaborations apply a lower cut on the photon
energy of about $1.8-2$~GeV, a  detailed knowledge of the photon
energy spectrum is mandatory in order to obtain a prediction for
the $\bar{B} \to X_s \gamma$ branching ratio.  The measured photon
energy spectrum provides direct information on the shape function,
while from the moments of the truncated spectrum it is possible to
extract relevant information on the Heavy Quark Expansion
parameters.

Working at NLO in $\alpha_s$, it is necessary to consider the
virtual corrections to the process $b \to s \gamma$ together with
the matrix elements of the corresponding process with a gluon in
the final state: $b \to s \gamma g$. At NNLO, it is also necessary
to consider the matrix elements with up to two gluons or a $q
\bar{q}$ pair in the final state.
After defining the dimensionless variable $z = 2 E_\gamma/m_b$ ($0\le z\le1$), the quantity
$G_{ij}$ in Eq.~(\ref{decay}) can be written as
\be
G_{ij} \left(E_0,m_b \right) = \int_{z_0}^1 \frac{d G_{ij} (z)}{dz} dz \, ,
\label{intt}
\ee
where one sets $z_0 = 2 E_0/m_b$ and $\mu = m_b$, and $d G_{ij}/dz$ is the photon energy spectrum.

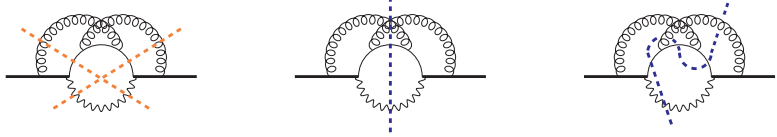
\begin{figure}[t]
\vspace*{.2cm} \hspace*{2.2cm}
\begin{tabular}{ccccc} \begin{picture}(0,0)(0,0)
\SetScale{.6}
  \SetWidth{1.8}
\Line(-60,0)(-20,0)
\Line(60,0)(20,0)
  \SetWidth{.5}
\CArc(0,0)(20,0,180)
\PhotonArc(0,0)(20,-180,0){2}{12.5}
%
%
\GlueArc(-13,11)(25,13.5,207){3}{15}
\GlueArc(13,11)(25,-27,166.5){3}{15}
%
%
  \SetWidth{1.8}
\SetColor{Orange}
\DashLine(-50,30)(30,-20){3}
\DashLine(50,30)(-30,-20){3}
\end{picture}
&
\hspace{3cm}
&
\begin{picture}(0,0)(0,0)
\SetScale{.6}
  \SetWidth{1.8}
\Line(-60,0)(-20,0)
\Line(60,0)(20,0)
  \SetWidth{.5}
\CArc(0,0)(20,0,180)
\PhotonArc(0,0)(20,-180,0){2}{12.5}
%
%
\GlueArc(-13,11)(25,13.5,207){3}{15}
\GlueArc(13,11)(25,-27,166.5){3}{15}
%
%
  \SetWidth{1.8}
\SetColor{Blue}
\DashLine(0,-35)(0,50){3}
\end{picture}
&
\hspace{3cm}
&
 \begin{picture}(0,0)(0,0)
\SetScale{.6}

  \SetWidth{1.8}
\Line(-60,0)(-20,0)
\Line(60,0)(20,0)
  \SetWidth{.5}
\CArc(0,0)(20,0,180)
\PhotonArc(0,0)(20,-180,0){2}{12.5}
%
%
\GlueArc(-13,11)(25,13.5,207){3}{15}
\GlueArc(13,11)(25,-27,166.5){3}{15}
%
%
  \SetWidth{1.8}
\SetColor{Blue}
\DashLine(-5,-30)(-20,15){3}
\DashCArc(-10,15)(10,8,55){3}
\DashCArc(-10,15)(10,115,172){3}
\DashCArc(10,15)(10,180,350){3}
\DashLine(24,27)(20,15){3}
\DashLine(28,39)(32,51){3}
\end{picture}
\end{tabular}\\
\caption{Cuts contributing to the $b \to s g$, $b \to gg$, and $b
\to s s \bar{s}$ processes in a three-loop $b$-quark self-energy
diagram. \protect\label{fig6}}
\end{figure}

The only component of the photon energy spectrum completely known
at NNLO in QCD is $G_{77}$
(Refs.~\refcite{Melnikov:2005bx,Asatrian:2006sm}), which
originates from the interference of graphs involving the magnetic
dipole operator $ O_7$. This component of the spectrum is the
numerically dominant one. The general structure of a spectrum
component $\OO{i}{j}$ is the following:
\be
\frac{d G_{ij}(z)}{dz} = f_{ij} \delta(1-z) + R_{ij}(z) \, . \label{strz}
\ee
Because of kinematic, only matrix elements with three or four
particles in the final state  contribute to the function $R(z)$.
Therefore, if the total decay rate $G_{ij}(z_0 = 0)$ is known,
 it is sufficient to
calculate the function $R(z)$ in Eq.~(\ref{strz}), keeping $z \neq 1$,
in
order to know the photon energy spectrum. The constant $f$ in Eq.~(\ref{strz}) can be fixed a posteriori
by requiring the equality of the total rate with the integral in
Eq.~(\ref{intt}) once one sets $z_0 = 0$. The $\OO{7}{7}$
component of the total decay rate can be found in
Refs.~\refcite{Blokland:2005uk} and \refcite{Asatrian:2006ph}.
 Consequently, in
order to calculate the corresponding spectrum component,  it is
sufficient to consider one-loop corrections to the process $b \to
s \gamma g$, as well as tree-level matrix elements contributing to
the processes $b \to s \gamma g g$ and $b \to s \gamma q \bar{q}$.
At NLO, there are just two Feynman diagrams contributing to the
$\OO{7}{7}$ spectrum for $z \neq 1$. The squared amplitudes of the
two graphs must be integrated over the final-state phase space,
and the integrand must be multiplied  by a Dirac delta function
fixing the energy of the photon. Schematically,
\vspace*{3mm} \be \left. \frac{d G_{77}(z)}{dz}
\right|_{\mbox{NLO}}\! \hspace*{-.2cm} \rightarrow \!\int
\mbox{phase space}\hspace*{1.8cm} \vcenter{\hbox{
  \begin{picture}(0,0)(0,0)
\SetOffset(-30,5)
\SetScale{.7}
\SetWidth{1.5}
\ArrowLine(-30,0)(0,0)
  \SetWidth{.5}
\SetColor{Black}
\ArrowLine(0,0)(30,0)
\Photon(0,0)(25,-25){3}{8.5}
\Gluon(-12,0)(25,25){3}{8}
\CCirc(0,0){3}{Black}{Blue}
\Line(-33,25)(-33,-25)
\end{picture}
+
\begin{picture}(0,0)(0,0)
\SetOffset(30,5)
\SetScale{.7}
\SetWidth{1.5}
\ArrowLine(-30,0)(0,0)
  \SetWidth{.5}
\SetColor{Black}
\ArrowLine(0,0)(30,0)
\Photon(0,0)(25,-25){3}{8.5}
\Gluon(12,0)(25,25){3}{5}
\CCirc(0,0){3}{Black}{Blue}
\Line(33,25)(33,-25)
\Text(29.5,12)[cb]{$2$}
\end{picture}
}}
\hspace*{1.8cm}
\times
 \delta \left(z-\frac{2 E_\gamma}{m_b}\right) \, .\label{opt1}
\ee
The optical theorem relates the decay width to the imaginary part
of two-loop $b$-quarks self-energy diagrams. The contribution of
each specific physical cut to the imaginary part of the two-loop
self-energy diagrams can be calculated by means of the Cutkosky
rules. In particular, the quantity in Eq.~(\ref{opt1}) is obtained
by summing over three different cut diagrams
\vspace*{4mm} \be \left. \frac{d G_{77}(z)}{dz}
\right|_{\mbox{NLO}}\! \hspace*{-.2cm}  \rightarrow
\!\hspace*{1.5cm} \vcenter{\hbox{\begin{picture}(0,0)(0,0)
\SetOffset(0,3) \SetScale{.6}
  \SetWidth{1.8}
\Line(-60,0)(-20,0)
\Line(60,0)(20,0)
  \SetWidth{.5}
\PhotonArc(0,0)(20,-180,0){2}{12.5}
\CArc(0,0)(20,0,180)
%
%
\GlueArc(0,0)(40,0,180){3}{19}
%
%
  \SetWidth{1.8}
\SetColor{Blue}
\DashLine(0,50)(0,-25){3}
\end{picture}
\hspace*{1.2cm}
+
\hspace*{1.2cm}
\begin{picture}(0,0)(0,0)
\SetOffset(0,3)
\SetScale{.6}
  \SetWidth{1.8}
\Line(-60,0)(-20,0)
\Line(60,0)(20,0)
  \SetWidth{.5}
\CArc(0,0)(20,0,180)
\PhotonArc(0,0)(20,-180,0){2}{12.5}
%
%
\GlueArc(0,20)(15,-158,-22){3}{7}
%
%
  \SetWidth{1.8}
\SetColor{Blue}
\DashLine(0,35)(0,-25){3}
\end{picture}
\hspace*{1.2cm}
+
\hspace*{1.2cm}
\begin{picture}(0,0)(0,0)
\SetOffset(0,3)
\SetScale{.6}
  \SetWidth{1.8}
\Line(-60,0)(0,0)
\Line(60,0)(40,0)
  \SetWidth{.5}
\CArc(20,0)(20,0,180)
\PhotonArc(20,0)(20,-180,0){2}{12.5}
%
%
\GlueArc(-5,10)(32,16.5,197){3}{16}
%
%
  \SetWidth{1.8}
\SetColor{Blue}
\DashLine(-13.5,55)(40,-30){3}
\end{picture}
}}
\hspace*{1.2cm}
\, ,\label{opt2}
\ee
where it is understood that the Dirac delta in Eq.~(\ref{opt1})
was inserted by hand in the integrands. The same principle can be
applied at NNLO, so that the $\OO{7}{7}$ photon energy spectrum
can be obtained by calculating appropriate cuts of three-loop
$b$-quark self-energy diagrams; examples  are shown in
Fig.~\ref{fig6}.
The contribution of each  cut to the photon energy spectrum can be
evaluated with the techniques discussed in the previous section.
The Dirac delta function employed in order to fix the photon energy can be
rewritten as a difference of propagators by means of  the identity in
Eq.~(\ref{cutprop})
already employed for the delta functions associated with cut propagators.
The MIs can then be calculated in different ways. The differential
equation method (Refs.~\refcite{de1,Argeri:2007up}) was employed
in both Ref.~\refcite{Melnikov:2005bx} and
Ref.~\refcite{Asatrian:2006sm}. In Ref.~\refcite{Asatrian:2006sm},
all of the MIs were checked numerically by means of the sector
decomposition method.
The sum of all of the cuts contributing to the magnetic dipole
component of the spectrum is free from collinear and soft
divergencies. In the intermediate stages of the calculation, the
latter are regulated by dimensional regularization. The residual
single pole in the dimensional regulator $\varepsilon$ is canceled
by ultraviolet renormalization.
%
The $\OO{7}{7}$ spectrum is singular in the $z \to 1$ limit; the
singularities appear in the spectrum as plus distributions of the
form $[\ln^n (1-z) / (1-z)]_+$ ($n = 0,\ldots,3$). The singular
terms were also predicted on the basis of the universality of soft
and collinear gluon radiation (Ref.~\refcite{Andersen:2005bj}).
%
It is interesting to observe that in the $z \to 0$ limit the
$\OO{7}{7}$ component of the spectrum should vanish like $z^3$.
This can be easily proved; each effective operator $O_7$ gives
raise to a factor $z$ in the spectrum, and an additional factor
$z$ originates from the integration over the photon phase space.
In a similar way, it is possible to determine the behavior of other
components of the spectrum in the $z \to 0$ limit.  Since the QED
coupling of the quarks with the photon is proportional to
$1/E_\gamma$ for soft photons, the $\OO{7}{8}$ component of the
photon energy spectrum is proportional to $z$ when $z \to 0$. In
the same limit, the $\OO{8}{8}$ component of the spectrum behaves like
$1/z$.
These kinds of considerations are based upon the Feynman rules
employed in the calculation; therefore, they apply to each cut of
each diagram contributing to a given spectrum component. The
qualitative information on the spectrum behavior in the $z \to 0$
limit can be used as a powerful check of the calculation, but it
can also be employed to fix some of the initial conditions needed
to calculate the MIs with the differential equation method.
%
%
In Refs.~\refcite{Blokland:2005uk} and \refcite{Asatrian:2006ph} the $c$-quark
mass is set to zero. The exact dependence of the NNLO corrections
to the $\OO{7}{7}$ spectrum and total decay rate on the $c$-quark
mass is found in Ref.~\refcite{Asatrian:2006rq}.

The techniques outlined above can in principle be used  to also  calculate
the NNLO QCD corrections to the $\OO{7}{8}$ component of the
spectrum. In particular, the dependence of the $\OO{7}{8}$
spectrum on the $c$-quark mass was studied in
Ref.~\refcite{Ewerth:2008nv}.

The BLM corrections  to all of the
components of the photon energy spectrum, with the
exception of $\OO{8}{8}$, were calculated almost a decade ago
(see Ref.~\refcite{Ligeti:1999ea}). It is well known that the BLM
corrections, which are proportional to $\alpha_s^2 \beta_0$,
usually provide a reliable estimate of the full  $\alpha_s^2$
corrections. In the case of the $\OO{7}{7}$ spectrum component it
was possible to verify that the BLM approximation provides
a good estimate of the complete NNLO corrections.
%
The BLM corrections to the $\OO{8}{8}$ spectrum component were not calculated
in Ref.~\refcite{Ligeti:1999ea}; they are suppressed by a factor $(Q_d
C_8/C_7)^2 \sim 0.03$, and therefore they are expected to be numerically small.
Moreover, the
$\OO{8}{8}$ spectrum component
includes logarithmic singularities in the limit in which the strange quark is
considered massless. As discussed above, the $\OO{8}{8}$ spectrum
is also singular in the limit of vanishing photon energy;
therefore, its contribution to the branching ratio becomes numerically large for
low (and experimentally unattainable) values of the cut on the photon energy.
For completeness, the BLM corrections to the $\OO{8}{8}$ spectrum were recently
computed in Ref.~\refcite{UliO8O8}; the collinear singularities were regulated by
keeping a finite strange quark mass $m_s$ when needed.

Finally, Table~\ref{Table1} summarizes the available NNLO QCD corrections to
the matrix elements of the $b \to \Xsp \gamma$ decay width, indicating
both calculations of the total decay rate and of the photon energy spectrum.

\begin{table}[ht]
\tbl{NNLO QCD corrections to the various component of the $b \to
\Xsp \gamma$ decay width.  The first row of the table refers to
the full $\alpha_s^2$ corrections, the second refers to the
$\alpha_s^2$ corrections with a light-quark loop, the third to
$\alpha_s^2$ corrections with a massive charm- or bottom-quark loop, and
the fourth to non-BLM corrections in the $m_c \gg m_b/2$
approximation. \label{Table1}} {\begin{tabular}{@{}cccccc@{}}
\toprule App. & $\OO{1}{1}$ & $\OO{1}{7}$ & $\OO{1}{8}$
    & $\OO{7}{7}$ & $\OO{7}{8}$\\
        & $\OO{1}{2}$ & $\OO{2}{7}$ & $\OO{2}{8}$ & & $\OO{8}{8}$ \\
    & $\OO{2}{2}$ & &  &  &  \\
\colrule
full $\alpha_s^2$ & & &  &  Refs.~\refcite{Blokland:2005uk,Asatrian:2006ph,Melnikov:2005bx,Asatrian:2006sm}&    \\
$\alpha_s^2 N_l$ &Ref.~\refcite{Ligeti:1999ea} &
Refs.~\refcite{Ligeti:1999ea,Bieri:2003ue,Boughezal:2007ny}& &
Refs.~\refcite{Ligeti:1999ea,Bieri:2003ue} &
Refs.~\refcite{Ligeti:1999ea,Bieri:2003ue,UliO8O8}   \\
$c$- $b$- loops & & Ref.~\refcite{Boughezal:2007ny}& &  Ref.~\refcite{Asatrian:2006rq}& Refs.~\refcite{Ewerth:2008nv,UliO8O8}   \\
$m_c \gg m_b/2$ &Ref.~\refcite{MisSte}  & Ref.~\refcite{MisSte} &
Ref.~\refcite{MisSte} & Ref.~\refcite{MisSte}  &    \\
\botrule
\end{tabular}}
\end{table}

\section{NNLO Estimate of the Branching Ratio}

According to the procedure of Ref.~\refcite{Gambino:2001ew}, which
is designed in order to reduce parametric uncertainties
originating from CKM angles and $c$- and $b$-quark masses, the
$\bar{B} \to X_s \gamma$ branching ratio can  be written as
\be{\mathcal B}\left( \bar{B} \to X_s \gamma \right)_{E > E_0} =
{\mathcal B}\left( \bar{B} \to X_c  e \bar{\nu}
\right)_{\mbox{exp}} \left|\frac{V^*_{ts} V_{tb}}{V_{cb}}
\right|^2 \frac{6 \alpha_{em}}{\pi C} \left[P(E_0) +N(E_0) \right]
\, , \label{BRth} \ee
where the perturbative corrections $P(E_0)$ are defined as follows
\be \left|\frac{V^*_{ts} V_{tb}}{V_{cb}} \right|^2\frac{6
\alpha_{em}}{\pi } P(E_0) = \left|\frac{V_{ub}}{V_{cb}}\right|^2
\frac{\Gamma\left(b \to \Xsp \gamma \right)_{E
>E_0}}{\Gamma(b \to X_u e \bar{\nu})} \, ,  \ee
 and $N(E_0)$ denotes non-perturbative corrections. The
factor $C$ is given by
\be C = \left|\frac{V_{ub}}{V_{cb}} \right|^2
\frac{\Gamma\left(\bar{B} \to X_c e \bar{\nu}
\right)}{\Gamma\left(\bar{B} \to X_u e \bar{\nu}\right)} \, . \ee
Since $C$ is a ratio of inclusive decay widths, it can be
calculated  by means of the same tools employed in the calculation
of the $\bar{B} \to X_s \gamma$ width, and expressed as a double
series in powers of $\alpha_s$ and of $\Lambda_{{\rm QCD}}/m_b$
(see Ref.~\refcite{Gambino:2008fj} for a recent analysis).

The current theoretical estimate of the quantity in
Eq.~(\ref{BRth}) in the SM  is (Refs.~\refcite{MisSte,manyaut})
\be {\mathcal B}\left(\bar{B} \to X_s \gamma\right)_{E > 1.6 \,
\mbox{{\tiny GeV}} }^{\mbox{{\small SM}}} = \, (3.15 \pm 0.23)
\times 10^{-4} \label{BRext} .\ee
The estimate in Eq.~(\ref{BRext}) includes all of the numerically
leading NNLO QCD corrections. When the value of $C$ obtained in
Ref.~\refcite{Gambino:2008fj} is employed, the central value of
Eq.~(\ref{BRext}) increases by a few percent.
The error on the theoretical estimate is  about $7 \%$, and  was
obtained by combining in quadrature four different uncertainties:
parametric uncertainty ($ 3 \%$), uncertainty due to missing
higher order corrections ($3 \%$), uncertainty due to
non-perturbative corrections ($5 \%$), and uncertainty due to the
$m_c$-interpolation ambiguity of Ref.~\refcite{MisSte} ($3 \%$).
The estimate of Eq.~(\ref{BRext}) does not include some NNLO and
non-perturbative corrections which are currently known; however,
the combined effect of the neglected contributions is about $+ 1.6
\%$, which is a small correction compared to the theoretical
uncertainty (more details can be found in Ref.~\refcite{Misiak:2008ss}).
It is important to observe that the result of
Refs.~\refcite{MisSte,manyaut} reduced the theoretical
uncertainty, which is  approximately of the same
magnitude as the experimental one. As expected, the inclusion of
NNLO corrections significantly reduced the dependence of the
branching ratio on the matching scale $\mu_\tw \sim M_\tw$, on the
low-energy scale $\mu_b \sim m_b$, and especially on the
charm-mass $\overline{{\small \mbox{MS}}}$ renormalization scale
$\mu_c$ that first enters the calculation at NLO.
The scale dependence of the LO, NLO, and NNLO predictions for the
branching ratio are compared in Fig.~2 of Ref~\refcite{manyaut}.
The SM theoretical prediction of Eq.~(\ref{BRext}) is now slightly
more than $1 \sigma$ lower than the experimental average in
Eq.~(\ref{WA}). Such an agreement can be used in order to set
stringent constraints on the parameters of some Beyond the SM
physics models (see for example Ref.~\refcite{manyaut}).

While progress in the calculation of perturbative corrections to
the $b \to \Xsp \gamma$ decay width is expected in the future, the
current theoretical error is dominated by the uncertainty
associated to non-perturbative effects, estimated to be about $5
\%$ (Ref.~\refcite{manyaut}). The non-perturbative uncertainty
primarily arises from corrections of order $O (\alpha_s
\Lambda_{{\rm QCD}} / m_b)$ which are very difficult to evaluate;
they were analyzed in Ref.~\refcite{PLN}.


In Table~\ref{Table2}, I summarize the size of NLO, NNLO, and
non-perturbative corrections to the branching ratio.
\begin{table}[ht]
\tbl{Size of various perturbative and non-perturbative set of
corrections to ${\mathcal B}(\bar{B} \to X_s \gamma)_{E
> 1.6 \mbox{{\tiny GeV}}}^{\mbox{{\tiny SM}}}$.  The percentages refer to
the size of a given set of corrections with respect to a LO
branching ratio of $\sim 3.4 \times 10^{-4}$. The corrections of
$O(\alpha_s \Lambda_{{\rm QCD}} /m_b)$ are not yet known.
\label{Table2}} {\begin{tabular}{@{}cc|cc@{}} \toprule
    Perturbative & &Non-perturbative & \\
\colrule
NLO QCD $O(\alpha_s)$ & $30 \%$ & LO QCD + NLO $m_b$ $O(\Lambda_{{\rm QCD}}^2 /m_b^2)$ & $1 \%$ \\
NLO EW $O(\alpha)$ & $4 \%$ & LO QCD + NLO $m_c$ $O(\Lambda_{{\rm QCD}}^2 /m_c^2)$& $3 \%$ \\
NNLO QCD $O(\alpha_s^2)$ & $10 \%$  & NLO QCD + NLO $m_b$ $O(\alpha_s \Lambda_{{\rm QCD}} /m_b)$& $5 \%$ \\
\botrule
\end{tabular}}
\end{table}

In conclusion, the program which aims to calculate  the NNLO QCD
corrections to the $b \to \Xsp \gamma$ decay is well under way.
The results so far obtained already have a substantial impact on
the theoretical prediction of the $\bar{B} \to X_s \gamma$ SM
branching ratio. The calculation of the NNLO corrections poses
numerous technical challenges related to the number of Feynman
diagram involved in various steps of the calculation, as well as
to the evaluation of the needed integrals. Such challenges require
the extensive application of the most current calculational
techniques developed for the automated evaluation of multi-loop
Feynman diagrams.

\appendix

\section{Effective Operators \label{Operators}}

For completeness, the eight effective operators
relevant for the $b \to \Xsp \gamma$ decay are listed below:

\bea O_1 &=& \left(\bar{s} \gamma_\mu T^a P_L c \right)
\left(\bar{c} \gamma^\mu T_a P_L b \right) \, , \nn \\
O_2 &=& \left(\bar{s} \gamma_\mu P_L c \right)
\left(\bar{c} \gamma^\mu  P_L b \right) \, , \nn \\
O_3 &=& \left(\bar{s} \gamma_\mu P_L b \right)
\sum_q \left(\bar{q} \gamma^\mu  q \right) \, , \nn \\
O_4 &=& \left(\bar{s} \gamma_\mu T^a P_L b \right)
\sum_q \left(\bar{q} \gamma^\mu T_a q \right) \, , \nn \\
O_5 &=& \left(\bar{s} \gamma_\mu \gamma_\nu \gamma_\rho
 P_L b \right)
\sum_q \left(\bar{q} \gamma^\mu \gamma^\nu \gamma^\rho q \right) \, , \nn \\
O_6 &=& \left(\bar{s} \gamma_\mu \gamma_\nu \gamma_\rho T^a
 P_L b \right)
\sum_q \left(\bar{q} \gamma^\mu \gamma^\nu \gamma^\rho T_a
q \right) \, , \nn \\
O_7 &=& \frac{e}{16 \pi^2} m_b
\left(\bar{s} \sigma^{\mu \nu} P_R b\right) F_{\mu\nu} \, , \nn \\
O_8 &=& \frac{g_s}{16 \pi^2} m_b \left(\bar{s} \sigma^{\mu \nu}
T^a P_R b\right) G_{\mu\nu} \, . \label{operators} \eea
In the above equations, $e$ and $F_{\mu\nu}$ ($g_s$ and
$G_{\mu \nu}$) represent the electromagnetic (strong) coupling constant
and field strength, respectively. The eight color generators are
indicated by $T^a$. The sums in the operators $O_3,\ldots,O_6$ run
over the light quarks.


\section*{Acknowledgments}

I would like to thank P.~Gambino for a number of discussions and
suggestions,  and  K.~Varade for a careful reading of the
manuscript. I am grateful to U.~Haisch for several interesting
discussions and for providing the numbers in Table~\ref{Table2}.
This work was supported by the Swiss National Science Foundation
(SNF) under contract 200020-117602.

%


\end{document}